\documentclass[journal]{IEEEtran}
\usepackage{graphicx}
\usepackage{caption}
\usepackage{pifont}
\usepackage{amssymb}%
\usepackage{amsmath}
\usepackage{hyperref}
\usepackage[utf8]{inputenc}
\usepackage[T1]{fontenc}
\usepackage{array}
\usepackage{longtable}
\usepackage{amssymb} % For \checkmark
\usepackage{pifont}  % For \times

\title{Eco-Friendly 0G Networks: Unlocking the Power of Backscatter Communications for a Greener Future}

\author{Shumaila Javaid \IEEEmembership{Member, IEEE}, Hamza Fahim \IEEEmembership{Member, IEEE}, Bin He \IEEEmembership{Senior Member, IEEE}, and Nasir Saeed \IEEEmembership{Senior Member, IEEE}
    \thanks{S. Javaid, H. Fahim, and B. He are with the Department of Control Science and Engineering, College of Electronics and Information Engineering, Tongji University, Shanghai 201804, and  National Key Laboratory of Autonomous Intelligent Unmanned  Systems, Tongji University, Shanghai 201210, China\\
    N. Saeed is with the Department of Electrical and Communication Engineering, College of Engineering, UAE University, Al-Ain 15551, UAE (e-mail: mr.nasir.saeed@ieee.org).}
}

\begin{document}

\maketitle

\begin{abstract}
Backscatter Communication (BackCom) technology has emerged as a promising paradigm for the Green Internet of Things (IoT) ecosystem, offering advantages such as low power consumption, cost-effectiveness, and ease of deployment. While traditional BackCom systems, such as RFID technology, have found widespread applications, the advent of ambient backscatter presents new opportunities for expanding applications and enhancing capabilities. Moreover, ongoing standardization efforts are actively focusing on BackCom technologies, positioning them as a potential solution to meet the near-zero power consumption and massive connectivity requirements of next-generation wireless systems. 0G networks have the potential to provide advanced solutions by leveraging BackCom technology to deliver ultra-low-power, ubiquitous connectivity for the expanding IoT ecosystem, supporting billions of devices with minimal energy consumption. This paper investigates the integration of BackCom and 0G networks to enhance the capabilities of traditional BackCom systems and enable Green IoT. We conduct an in-depth analysis of BackCom-enabled 0G networks, exploring their architecture and operational objectives, and also explore the Waste Factor (WF) metric for evaluating energy efficiency and minimizing energy waste within integrated systems. By examining both structural and operational aspects, we demonstrate how this synergy enhances the performance, scalability, and sustainability of next-generation wireless networks. Moreover, we highlight possible applications, open challenges, and future directions, offering valuable insights for guiding future research and practical implementations aimed at achieving large-scale, sustainable IoT deployments.
\end{abstract}


\section{Introduction}
The Internet of Things (IoT) has rapidly emerged as one of the most transformative technologies of the 21st century, enabling seamless connectivity between individuals, devices, and sensors within intelligent environments. This ongoing advancement across sectors such as automation, data acquisition, and real-time communication has significantly enhanced operational efficiency and user experience \cite{laghari2021review,villamil2020overview}. However, as IoT continues to expand, it presents significant environmental challenges, particularly in terms of the sustainability of conventional systems and infrastructures that were not designed to support the scale of interconnected devices. The increasing number of connected devices has raised concerns about higher energy consumption and the accumulation of electronic waste, making the sustainability of these systems an urgent issue. In response to these challenges, the concept of green IoT has gained significant attention, advocating for the adoption of energy-efficient solutions and sustainable practices in powering IoT devices \cite{albreem2021green,maksimovic2018greening}.

Backscatter Communication (BackCom) has gained considerable importance as a solution for achieving greener IoT systems by offering energy efficiency and cost-effectiveness \cite{niu2019overview,liu2019next}. This technology gained widespread prominence in the early 2000s with the advent of Radio Frequency Identification (RFID), a key BackCom application that revolutionized industries, including retail, logistics, and supply chain management by enabling wireless identification and tracking of objects \cite{yao2020backscatter,couraud2017real}. BackCom enables BackCom Devices (BDs) to harvest energy from surrounding Radio Frequency (RF) signals to eliminate the need for limited-life batteries and significantly reduce maintenance requirements. In addition, BDs transmit signals by modulating their impedance to reflect the incoming RF signals, mitigating the need for active components such as mixers, digital-to-analog converters, and RF synthesizers. This approach improves energy efficiency while simultaneously simplifying the overall design of the system \cite{zhang2019robust,li2018hybrid,zhao2018ambient}. %In parallel, academic research has increasingly focused on ambient backscatter techniques, which use existing RF signals for communication. This approach further improves the energy efficiency and sustainability of BackCom systems by improving data transmission rates, extending communication ranges, and integrating advanced energy harvesting mechanisms .

This growing focus has led to the integration of multi-antenna systems and advanced modulation schemes, marking pivotal advancements in BackCom technology aimed at achieving higher data throughput and more reliable communication channels. Furthermore, the convergence of BackCom with energy harvesting techniques such as photovoltaic cells and thermal energy converters has reduced dependence on traditional power sources, enhancing the sustainability of IoT deployments. These advances address the energy constraints inherent in IoT devices, enabling the development of large-scale, maintenance-free sensor networks capable of supporting a wide range of IoT applications, from smart agriculture and healthcare to industrial automation and smart cities \cite{liu2023covert,duan2018multi}.

In parallel, 0G networks represent another frontier that can take advantage of advances in BackCom technology to improve their performance and efficiency. Designed to provide ultra-low-power, ubiquitous connectivity for the expanding IoT ecosystem, 0G networks aim to support billions of devices with minimal energy consumption, overcoming the limitations of traditional active communication methods that demand significant power \cite{mehta20140g,mondal2015survey}. BackCom, with its ability to operate through passive signal reflection, is ideally suited for this purpose, enabling 0G networks to function with dramatically lower energy overhead. This integration enhances the scalability and sustainability of IoT deployments and addresses key challenges related to energy consumption and infrastructure costs. Therefore, investigating the convergence of BackCom and 0G networks is imperative for the development of the next generation of wireless communication systems that are both environmentally sustainable and capable of supporting large-scale IoT applications.

This paper acknowledges the transformative potential of integrating BackCom systems with emerging 0G networks, emphasizing how this collaboration can drive the creation of sustainable and scalable IoT ecosystems. By thoroughly examining critical challenges such as energy efficiency, scalability, and infrastructure dependency, this paper aims to identify the limitations of current IoT solutions and offers valuable insights into overcoming these barriers. In addition to conventional performance metrics such as energy efficiency and scalability, this work introduces the Waste Factor (WF) as a novel metric to evaluate the environmental impact of communication technologies \cite{10604822}. The WF quantifies the proportion of energy expended during transmission and processing that does not directly contribute to successful communication, such as dissipated power and redundant operations. By incorporating this metric, the study highlights inefficiencies often overlooked by traditional evaluations, offering a more holistic perspective on a system's environmental footprint. Evaluating the WF as a critical parameter in the design of BackCom-Enabled next-generation communication systems aims to provide deeper insights into resource optimization, driving innovation toward greener and more sustainable wireless solutions. Through this comprehensive analysis, this work seeks to enhance the design and implementation of green IoT systems that are both environmentally sustainable and capable of supporting large-scale deployments. The key contributions of this study are highlighted as follows:

\begin{itemize}
    \item First, we provide a comprehensive overview of BackCom technologies, detailing their primary components, fundamental principles, recent advancements, and pivotal roles in advancing sustainable IoT frameworks. This extensive background establishes a solid foundation for understanding the capabilities and limitations of BackCom systems, emphasizing their significance within green IoT initiatives.

    \item Second, we conduct an in-depth analysis of 0G networks and their integration with BackCom systems, exploring their architecture and objectives for achieving ultra-low power and ubiquitous connectivity. This analysis includes an evaluation of the waste factor that assesses the efficiency of energy utilization within the integrated systems. By examining both the structural and operational aspects, we demonstrate how seamlessly integrating BackCom into the 0G framework enhances the performance and potential of next-generation wireless networks while minimizing energy waste to promote more sustainable and efficient IoT ecosystems.

    \item Third, we explore a range of synergistic applications resulting from the combined use of BackCom and 0G networks across various domains. By presenting detailed case studies in areas such as smart agriculture, healthcare monitoring, and urban infrastructure management, we illustrate how this integrated framework addresses real-world IoT challenges. These applications highlight the practical benefits and transformative potential of leveraging BackCom within 0G environments to foster resilient and eco-friendly smart ecosystems.

    \item Finally, we identify and examine the key challenges in deploying BackCom-enabled 0G networks, such as energy harvesting, interference management, spectrum utilization, and integration with emerging technologies. We highlight open challenges and outline future directions to ensure the robustness and scalability of integrated BackCom and 0G systems. This critical evaluation offers valuable insights for guiding future research and practical implementations aimed at achieving large-scale, sustainable IoT deployments.
\end{itemize}

\section{Backscatter Communication (BackCom)}
In this section, we present a comprehensive overview of BackCom systems, focusing on their primary classifications and techniques. We explore various types of BackCom systems, examining their contributions to energy efficiency and sustainability, and highlighting their significance for enabling green IoT solutions.

\subsection{Overview of BackCom Systems}
BackCom is a wireless communication paradigm that enables ultra-low-power data transmission by reflecting and modulating incident RF signals instead of generating new ones. This technique allows BDs (also known as tags) to communicate by altering the properties of incoming RF signals, such as amplitude, phase, or frequency, without the need for power-intensive active transmitters. A typical BackCom system comprises three main components, including the RF source, the BD, and the receiver. The RF source emits continuous or intermittent RF signals, which can be dedicated emitters such as RFID readers or ambient RF signals from existing wireless infrastructure such as television towers, Wi-Fi access points, and cellular Base Stations (BSs). The BD is a low-power or passive entity equipped with an antenna and a simple modulation circuit. It modulates its antenna impedance to reflect and encode data onto the incident RF signal. The receiver then detects and demodulates the backscattered signals to recover the transmitted information, requiring high sensitivity to discern the weak backscattered signals from ambient RF noise. The operational mechanism of BackCom involves energy harvesting, impedance modulation, signal reflection, and data reception. BDs harvest energy from incident RF signals or other ambient sources (e.g., solar or thermal energy) to power their circuitry, enabling battery-free operation. By changing the impedance of their antennas, these devices alter the reflection coefficient, thus modulating the backscattered signal using schemes such as On-Off Keying (OOK), Frequency-Shift Keying (FSK), or Phase-Shift Keying (PSK). The modulated signal is reflected back toward the receiver or propagates through the environment if ambient RF sources are used. The receiver captures the backscattered signal and employs signal processing techniques to extract the data from the weak and potentially noisy reflections \cite{ji2019efficient,van2018ambient}.

\begin{figure*}[h]
\centering
 \includegraphics[width=18cm]{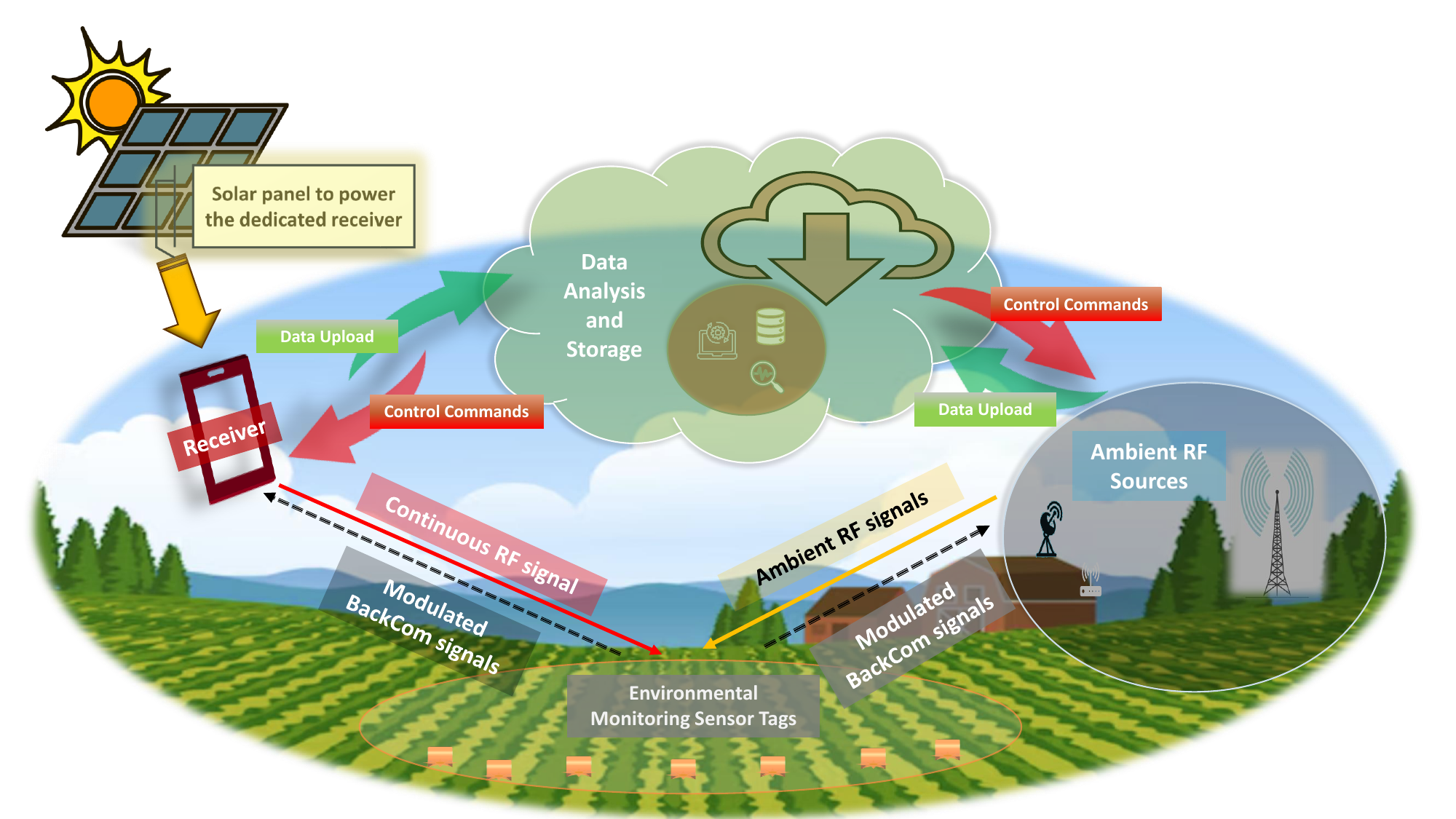}
  \caption{Hybrid Backscatter Communication System for Smart Agriculture.}
  \label{fig:backcom}
  \end{figure*}

Figure \ref{fig:backcom} illustrates a hybrid BackCom system that integrates both dedicated RF sources and ambient RF sources. It demonstrates how the system can utilize a dedicated RF emitter (e.g., an RFID reader) to power the BDs directly while also leveraging ambient RF sources such as Wi-Fi, cellular towers, or Long Range (LoRa) signals to provide additional energy and communication channels. The backscatter tags modulate the incoming RF signals from both dedicated and ambient sources and reflect them back as modulated signals, which are then captured by receivers. This hybrid approach enhances the system's energy efficiency and extends the communication range by utilizing multiple RF sources, ensuring robust data collection and transmission in environments such as smart agriculture.

\subsection{BackCom Systems Primary Classification}
BackCom systems can be classified based on their operational mechanisms and system configurations. The following sections provide detailed explanations, and Table \ref{tab:backcomsystems}
presents a comparative analysis of various BackCom platforms, highlighting their strengths and limitations across key performance criteria.

\subsubsection{Ambient Backscatter} It leverages existing RF signals in the environment, such as those from TV towers, Wi-Fi routers, or cellular BSs, enabling BDs to communicate without the need for dedicated RF sources. This capability makes ambient backscatter particularly well-suited for ultra-low-power IoT applications, including smart agriculture and environmental monitoring, where deploying dedicated transmitters may be impractical \cite{wu2022survey}. Numerous studies \cite{zhao2018ambient,darsena2017modeling,ma2018blind} have investigated the channel capacity and outage performance of ambient backscatter systems, demonstrating how different types of RF signals, such as those from Wi-Fi, cellular, and TV sources, affect system performance. These investigations have revealed significant variations in channel capacity based on factors such as signal frequency, power density, and modulation schemes employed by different RF sources. Existing studies have also investigated multitag ambient backscatter systems, where multiple tags harvest RF energy and backscatter signals for data transmission. In particular, optimization approaches have been introduced in the literature to maximize user rates by jointly optimizing time allocation and power reflection, addressing Channel State Information (CSI) mismatches \cite{li2018hybrid,zhang2019robust}. These advancements underscore the versatility, technical challenges, and transformative potential of ambient backscatter communication, establishing it as a pivotal technology for future scalable, energy-efficient IoT networks.

\subsubsection{RFID-based Backscatter}It is one of the most prevalent forms of backscatter technology, in which a dedicated reader transmits RF signals to passive RFID tags. These tags, equipped with minimal circuitry, modulate their load impedance to reflect the incoming signals back to the reader, thereby embedding data such as unique identifiers or sensor information within the reflected signal. This bidirectional communication enables efficient identification and tracking of objects, making RFID-based backscatter indispensable in retail, supply chain management, and inventory tracking applications \cite{salvati2023emerging}. Recent studies \cite{jeong2020machine,alfian2020improving} have applied Machine Learning (ML) techniques to optimize RFID systems, demonstrating improvements in read range, data throughput, and scalability, particularly in dense tag environments. The authors in \cite{dastres2022radio,mao2021energy} explored energy-efficient power management strategies to extend the operational lifetime of passive RFID tags, facilitating large-scale deployments with minimal maintenance. Moreover, innovations in antenna design, particularly the development of Ultra-High Frequency (UHF) RFID systems, have been extensively investigated to increase the range and capacity of RFID-based BackCom systems \cite{dastres2022radio,abdulghafor2021recent}. Existing studies \cite{munoz2021rfid,abdulghani2022analysis,long2019full} have also focused on designing security protocols specifically designed for RFID systems to mitigate the risks associated with unauthorized access and data interception to address the critical vulnerabilities in passive communication systems. These developments underscore the growing significance of RFID-based backscatter technology in advancing next-generation IoT ecosystems, ensuring reliable object tracking, optimized energy usage, and strengthened security for large-scale, efficient deployments.

\subsubsection{Monostatic Backscatter} In this configuration, the transmitter and receiver are co-located (i.e., typically integrated within the same device). The transmitted RF signal is reflected off the backscatter tag and subsequently received by the same antenna or system that emitted the signal. This seamless interaction facilitates efficient and real-time data exchange, making monostatic backscatter ideal for applications requiring close-range identification and rapid response, such as proximity detection systems and security access control systems \cite{he2020monostatic}. Extensive research has been conducted to optimize the performance and reliability of monostatic backscatter systems to develop advanced signal processing algorithms that enhance detection accuracy and reduce interference in dense tag environments, significantly improving system robustness. The authors in \cite{wang2021energy,zhang2024design} explored energy-efficient modulation schemes tailored for monostatic setups, demonstrating substantial improvements in power consumption without compromising data integrity. Innovations in antenna design, such as the integration of Multiple-Input Multiple-Output (MIMO) technologies, have also been investigated to increase the range and data throughput of monostatic backscatter systems \cite{al2020massive}. Furthermore, recent studies \cite{gu2024breaking,abdulghani2022analysis} have also addressed security vulnerabilities inherent in monostatic configurations by implementing robust encryption protocols to ensure secure data transmission and prevent unauthorized access. These advancements underscore the versatility and technical prowess of monostatic BackCom, positioning it as a pivotal technology for enabling scalable, secure, and efficient IoT networks in environments where close-range communication is essential.

\subsubsection{Bistatic Backscatter} It involves the transmitter sending RF signals to the BD, while a remote receiver captures the signals reflected back by the BD. This spatial separation enables bistatic backscatter to excel in scenarios requiring long-range communication and minimal interference between the transmission and reception processes. Bistatic backscatter is particularly useful in applications such as remote sensing, wildlife monitoring, and environmental data collection, where readers and tags operate over extended distances and in diverse environmental conditions \cite{gu2024breaking,yao2020backscatter}. Existing studies \cite{rezaei2020large,rezaei2023coding} have developed advanced signal processing algorithms that improve the detection accuracy of reflected signals in noisy and dynamic environments, significantly improving the robustness of the system in remote sensing applications. The studies \cite{wang2021energy,yang2021energy} explored energy-efficient modulation schemes tailored for bistatic setups, demonstrating substantial reductions in power consumption without compromising data integrity. Recent advancements in antenna design, particularly the development of high-gain directional antennas, have been extensively studied to extend the effective range and improve the signal strength of bistatic backscatter systems. Using adaptive beamforming and diversity techniques, these advances effectively mitigate challenges such as multipath fading and signal attenuation in outdoor environments, thus significantly enhancing the reliability of data transmission in wildlife monitoring applications \cite{jia2021intelligent,tao2021novel,he2020monostatic}.

\subsubsection{Full-Duplex Backscatter} It enables devices to transmit and receive signals simultaneously, significantly reducing communication latency and enhancing data throughput. Unlike half-duplex systems, where transmission and reception occur at separate times or frequencies, full-duplex backscatter allows for real-time, bidirectional data exchange within the same frequency band. This is significantly beneficial for high-speed communication applications, such as next-generation IoT devices and wireless sensor networks, where rapid and continuous data exchange is essential for functionalities such as real-time monitoring, autonomous control, and dynamic data analytics. Existing studies have focused on optimizing full-duplex backscatter systems, addressing challenges such as self-interference cancelation and signal synchronization. The authors in \cite{zhang2019adaptive} developed advanced self-interference cancellation techniques that mitigate the interference caused by simultaneous transmission and reception to improve signal clarity and system reliability. Additionally, recent studies \cite{abdallah2023channel,liu2017full} have also focused on the implementation of adaptive and directional antennas to optimize signal propagation and reception in full-duplex environments. Furthermore, the authors in \cite{jafari2021maximizing,liu2014enabling} have also focused on enhancing the energy efficiency of full-duplex backscatter systems by exploring energy-harvesting mechanisms. These approaches enable continuous, self-sustained operation, significantly reducing the reliance on frequent battery replacements. These advancements underscore the transformative potential of full-duplex BackCom in enabling scalable, high-performance IoT networks. However, challenges remain, including the need for more efficient interference management, robust security protocols to protect simultaneous data streams, and the development of standardized frameworks to ensure interoperability across diverse IoT platforms.

\begin{table*}[h]
    \centering
     \caption{Comparison of Different Backscatter Communication Platforms Based on Key Performance Criteria.}
    \label{tab:backscatter_comparison}
    \begin{tabular}{|l|c|c|c|c|c|c|c|c|}
    \hline
    \textbf{Types} & \textbf{\begin{tabular}[c]{@{}c@{}}Low Power\\ Consumption\end{tabular}} & \textbf{\begin{tabular}[c]{@{}c@{}}High Data\\ Rate\end{tabular}} & \textbf{\begin{tabular}[c]{@{}c@{}}Wide\\ Communication\\ Range\end{tabular}} & \textbf{\begin{tabular}[c]{@{}c@{}}Interference\\ Resistance\end{tabular}} & \textbf{\begin{tabular}[c]{@{}c@{}}Simple\\ Implementation\end{tabular}} & \textbf{\begin{tabular}[c]{@{}c@{}}Energy\\ Harvesting\\ Capability\end{tabular}} & \textbf{\begin{tabular}[c]{@{}c@{}}Supports\\ Multiple\\ Antennas\end{tabular}} & \textbf{\begin{tabular}[c]{@{}c@{}}Adaptive\\ Modulation\end{tabular}} \\ \hline
    Ambient Backscatter & $\checkmark$ & & $\checkmark$ & $\checkmark$ & $\checkmark$ & & & \\ \hline
    RFID-based Backscatter & $\checkmark$ & & & $\checkmark$ & $\checkmark$ & $\checkmark$ & & \\ \hline
    Monostatic Backscatter & & $\checkmark$ & & $\checkmark$ & $\checkmark$ & & & \\ \hline
    Bistatic Backscatter & & $\checkmark$ & & $\checkmark$ & $\checkmark$ & & & \\ \hline
    Full-Duplex Backscatter & $\checkmark$ & $\checkmark$ & & $\checkmark$ & & & & \\ \hline
    Frequency-Shifted Backscatter & & $\checkmark$ & $\checkmark$ & & $\checkmark$ & & $\checkmark$ & \\ \hline
    Modulated Backscatter & & $\checkmark$ & $\checkmark$ & $\checkmark$ & $\checkmark$ & & & $\checkmark$ \\ \hline
    \end{tabular}
\end{table*}
\subsubsection{Frequency-Shifted Backscatter} It is an advanced backscatter technique that involves shifting the frequency of the reflected signal, thus enabling better separation from the original transmitted signal. This frequency shift mitigates interference, making frequency-shifted backscatter particularly effective in dense RF environments such as urban areas with a high concentration of Wi-Fi, cellular, and Bluetooth signals. By operating on a different frequency band, this technique ensures more robust and reliable communication in crowded spectrums, addressing the challenges posed by overlapping and congested RF channels. Extensive research \cite{ding2020harmonic,hessar2019netscatter} has been conducted to optimize frequency-shifted backscatter systems, focusing on enhancing signal separation, improving energy efficiency, and increasing data throughput. In \cite{van2018ambient,toro2022backscatter}, the authors studied frequency-shifting algorithms that dynamically adjust the reflection frequency based on ambient signal conditions, significantly reducing interference and enhancing signal clarity in real-time. The authors in \cite{vougioukas2018switching} explored the integration of frequency-shifted backscatter with cognitive radio technologies, enabling adaptive spectrum access and further mitigating interference in highly dynamic RF environments. Furthermore, Zhang et al. in \cite{zhang2016enabling} introduced frequency-shifted backscatter for ultra-low power on-body sensors using existing radios on smartphones and wearables. By shifting the backscattered signal to an adjacent frequency band, this technique reduces interference, enabling more robust decoding and efficient communication while consuming minimal power, making it practical for on-body sensing. These advancements highlight the versatility of frequency-shifted backscatter in optimizing communication efficiency in dense RF environments and enabling ultra-low power applications. As research continues to refine these techniques, frequency-shifted backscatter is poised to play a critical role in next-generation IoT systems, particularly where energy efficiency and reliable data transmission are essential in challenging spectral conditions.

\subsubsection{Modulated Backscatter}Modulated backscatter involves encoding data onto reflected RF signals by dynamically altering the impedance of backscatter devices (BDs). This modulation process can be implemented using various schemes, such as Amplitude Shift Keying (ASK), Phase Shift Keying (PSK), Frequency Shift Keying (FSK), and Pulse Width Modulation (PWM). Each of these techniques offers unique advantages and trade-offs in terms of data rate, energy consumption, and system complexity \cite{yang2017modulation}. For example, ASK is favored for its simplicity and low power requirements, making it suitable for low-data-rate sensor applications in IoT environments. Conversely, PSK and FSK provide higher data rates and greater resilience to noise, which are critical for more complex systems like wireless sensor networks and real-time monitoring. Meanwhile, PWM modulates the width of pulses to balance simplicity and data throughput, enabling flexibility across diverse IoT scenarios \cite{correia2017quadrature, qian2018iot}. Despite their widespread use, traditional modulation schemes have several limitations. Energy-efficient methods such as ASK and PWM suffer from low data throughput and high sensitivity to noise and interference, making them unsuitable for high-speed communication in crowded RF environments. Additionally, these methods exhibit low spectral efficiency, which is a significant drawback in overlapping frequency channels common in urban IoT deployments. While PSK and FSK achieve higher data rates, they require more complex circuitry and higher power consumption, making them impractical for energy-constrained devices. To address these limitations, more advanced modulation techniques have been developed. For instance, the authors in \cite{thomas2012quadrature} introduced strategies that leverage complex-valued backscatter to enable the transmission of multiple bits per symbol, significantly boosting data rates without increasing power consumption. This approach allows backscatter systems to achieve higher data throughput while maintaining energy efficiency, making them more suitable for next-generation IoT applications. Additionally, recent studies \cite{goudeli2020spatial} have explored integrating BackCom into 5G IoT networks using Spatial Modulation (SM) techniques with multiple antennas. By utilizing generalized Spatial Modulation (GSM), these systems exploit antenna indices to enhance spectral efficiency and reduce Symbol Error Rates (SER) to improve communication reliability in dense, high-interference environments.

\section{State-of-the-Art Backscatter Platforms}
In this section, we present a comprehensive overview of state-of-the-art ambient backscatter communication (BackCom) platforms, emphasizing their evolution and influence on greet IoT systems. Table \ref{tbl_comparison} highlights a comparative analysis of these platforms in various key parameters.

% Please add the following required packages to your document preamble:

\begin{table*}[]
\fontsize{18pt}{18pt}\selectfont
\renewcommand{\arraystretch}{1.8} % Adjust the
\caption{Comparison of different backscatter platforms.}
\label{tbl_comparison}
\large
\renewcommand{\arraystretch}{1.0} % Adjust the number as needed (default is 1)
\resizebox{\textwidth}{!}{%
\begin{tabular}{|l|l|l|l|l|}
\hline
\textbf{Feature} &
  \textbf{LoRa-Based Backscatter} &
  \textbf{Bluetooth Backscatter} &
  \textbf{Wi-Fi Backscatter} &
   \textbf{Cellular Backscatter} \\ \hline
\textbf{Range} &
  Up to 15 km (rural) &
  Up to 100 meters &
  Up to 100 meters &
  Up to 10 km (urban areas) \\ \hline
\textbf{Data Rate} &
  0.3 to 50 kbps &
  Up to 2 Mbps &
  Up to 1 Gbps &
  10 kbps to 1 Mbps\\ \hline
\textbf{\begin{tabular}[c]{@{}l@{}}Power \\ Consumption\end{tabular}} &
  Very Low &
 Moderate &
  High &
  Low to Moderate \\ \hline
\textbf{Scalability} &
  \begin{tabular}[c]{@{}l@{}}Highly Scalable \\ (millions of devices)\end{tabular} &
  Moderate &
  Highly Scalable &
 \begin{tabular}[c]{@{}l@{}}Scalable \\ (depending on infrastructure) \end{tabular}   \\ \hline
\textbf{Cost} &
  \begin{tabular}[c]{@{}l@{}}Low (leverages existing \\ LoRa gateways)\end{tabular} &
  Low to Moderate &
  High &
  Moderate to High \\ \hline
\textbf{Use Cases} &
  \begin{tabular}[c]{@{}l@{}}IoT, Smart Cities, \\ Agriculture\end{tabular} &
  \begin{tabular}[c]{@{}l@{}}Wearables, Home \\ Automation\end{tabular} &
  \begin{tabular}[c]{@{}l@{}}High-Speed Data \\ Transfer, Streaming\end{tabular} &
  \begin{tabular}[c]{@{}l@{}}Asset Tracking, Smart \\ Transportation\end{tabular} \\ \hline
\end{tabular}%
}
\end{table*}

\subsection{LoRa BackComs}
LoRa is a long-range, low-power wireless communication technology with chirp spread spectrum (CSS) modulation. It offers a data rate from 0.3 to 50 kbps and a transmission power of up to 20 dBm, supporting communication ranges up to 15 km \cite{guo2021efficient,jiang2021long}. This makes LoRa backscatter an attractive option for IoT applications requiring extensive coverage and energy efficiency, such as smart agriculture, environmental monitoring, and remote asset tracking. By integrating backscatter techniques with LoRa, BDs can transmit data by reflecting existing LoRa signals, thereby eliminating the need for active RF sources and significantly reducing power consumption. Recent studies \cite{jiang2021long, guo2021efficient, lazaro2021room, biswas2021direct,xiao2022backscatter} have focused on optimizing LoRa backscatter systems by improving signal reliability, energy harvesting efficiency, and data throughput. Moreover, advanced signal processing algorithms have been developed to mitigate interference from multiple BDs, enhancing the scalability and performance of LoRa networks in densely populated areas. These advancements have also addressed key challenges such as synchronization and latency in LoRa BackComs, making the technology more efficient and suitable for large-scale IoT applications. Moreover, existing studies \cite{talla2017lora,lin2023lora,katanbaf2021simplifying} have also focused on optimizing antenna design, such as the use of directional antennas and adaptive impedance matching circuits to extend the communication range and improve the overall performance of LoRa backscatter systems. However, challenges such as limited data rates compared to active LoRa systems, susceptibility to environmental interference, and the need for standardized backscatter protocols persist, driving ongoing research efforts aimed at enhancing the robustness and versatility of LoRa backscatter technologies.

\subsection{BLE BackComs}
It leverages the capabilities of Bluetooth Low Energy (BLE), a version of Bluetooth specifically designed for IoT devices \cite{zhang2020reliable}. BLE BC utilizes ambient BLE signals by reflecting and modulating these signals to transmit data, eliminating the need for dedicated power sources or active transmitters. This passive communication mechanism significantly reduces energy consumption and deployment costs by capitalizing on the widespread availability of BLE networks in residential, commercial, and public spaces. It offers data rates ranging from 125 kbps to 2 Mbps, providing a balance between high-speed data transfer and energy efficiency \cite{park2020adaptable,di2017performance}. This range makes it suitable for a variety of IoT applications, from low-data-rate sensor readings to more sophisticated systems requiring higher data throughput. Additionally, BLE Backscatter supports communication ranges of up to approximately 100 meters in open environments. However, this can vary based on factors such as device power, antenna design, and environmental conditions, ensuring reliable connectivity in diverse settings and enhancing the versatility of BLE Backscatter in both indoor and outdoor IoT deployments \cite{ensworth2017ble}. Extensive research has been conducted to optimize BLE backscatter systems, focusing on improving data throughput, minimizing interference, and enhancing the reliability of signal reflection. For instance, the authors in \cite{rosenthal2020dual} developed a dual-band, dual-mode backscatter approach that enhances data rates while maintaining ultra-low power consumption. By using time-division multiplexing and shared hardware, they improved energy efficiency by over 50 times compared to standard BLE transmitters, making BLE backscatter more suitable for high-demand environments. Additionally, recent studies \cite{ luo2024accurate,zhang2020enabling, chen2021reliable,memon2020ambient} have explored adaptive power management strategies that dynamically adjust the reflection coefficients of BDs based on ambient signal strength, significantly extending the operational lifetime of IoT devices. Existing studies have explored the implementation of miniaturized and multi-band antennas to improve signal reception and transmission efficiency in BLE backscatter systems. In parallel, researchers have addressed challenges such as signal interference and multipath fading in dense urban environments by integrating intelligent signal processing algorithms. These efforts have significantly enhanced the robustness and reliability of BLE BackComs, leading to more stable performance in high-density settings, making the systems better suited for complex and urban IoT applications \cite{wang2024enabling, song2021advances}. Despite recent advancements, several challenges persist in BLE backscatter systems, including limited communication range, susceptibility to environmental noise, and the lack of standardized protocols. Addressing these issues remains a critical focus of ongoing research aimed at enhancing the system's performance, reliability, and flexibility. Researchers are actively working to overcome these limitations by improving communication robustness and developing universal protocols, ultimately aiming to expand the usability and effectiveness of BLE backscatter in diverse and demanding application scenarios.

\subsection{Wi-Fi BackComs}
Wi-Fi BackCom techniques capitalize on the widespread availability of Wi-Fi signals and devices and have become a significant area of backscatter implementation. By utilizing existing Wi-Fi infrastructure, BDs can communicate by reflecting and modulating ambient Wi-Fi signals without the need for dedicated power sources or transmitters. This approach reduces both energy consumption and deployment costs by leveraging the widespread availability of Wi-Fi networks in residential, commercial, and public spaces \cite{abedi2020witag}. Wi-Fi backscatter offers data rates ranging from 0.3 to 50 kbps, balancing the need for sufficient data throughput for IoT applications with the energy efficiency characteristic of passive communication systems. While standard Wi-Fi technologies support much higher data rates, backscatter implementations achieve these lower rates due to the passive nature of BackCom, which limits both the modulation complexity and achievable data throughput. In addition, Wi-Fi backscatter can support communication ranges of up to 100 meters in open environments \cite{zhao2018x,bharadia2015backfi}. However, various factors, such as device power levels, antenna design, and environmental conditions, can significantly influence this range. Wi-Fi backscatter is particularly advantageous for IoT applications such as smart home devices, asset tracking, and environmental monitoring, where continuous and low-power communication is essential \cite{zhang2021commodity}.

Existing studies \cite{zhao2018spatial, abedi2020witag, bharadia2015backfi} have focused on optimizing Wi-Fi backscatter systems to enhance data throughput, extend communication range, and improve energy harvesting efficiency. The state-of-the-art literature \cite{kwon2021scalable, xu2018practical} has investigated energy harvesting and battery-free operation, enabling BDs to be powered solely from ambient wireless signals to eliminate the need for dedicated power sources and enhancing the sustainability of IoT deployments. Recent studies have also addressed interference and reliability challenges by mitigating interference from ambient signals and improving the reliability of Wi-Fi BackCom systems in real-world environments through enhanced synchronization and robust modulation techniques. Moreover, the integration of Wi-Fi backscatter systems into mobile and wearable devices has been explored to optimize energy consumption based on communication needs. For example, authors in \cite{wang2020low, zhao2019ofdma} have developed advanced modulation schemes that increase data rates by dynamically adjusting reflection coefficients based on real-time signal conditions to optimize the efficiency of Wi-Fi backscatter systems. Additionally, the existing focus on antenna design has played a crucial role in advancing Wi-Fi BackCom, particularly through reconfigurable antennas that adapt to various Wi-Fi channels and frequencies. This enhances the flexibility and reliability of backscatter systems in diverse operational settings, addressing challenges such as multipath fading and signal interference in dense Wi-Fi environments and significantly improving both signal clarity and data integrity \cite{kim2018exploiting,chen2022interference, kellogg2017passive}. To further enhance the performance of Wi-Fi backscatter systems, there is a need to explore hybrid communication models that integrate it with other low-power technologies to optimize its  performance across diverse IoT applications.

\subsection{Cellular BackComs}
Cellular BackCom uses existing cellular signals, particularly Long-Term Evolution (LTE) signals, as a continuous and robust RF source to facilitate passive data transmission. BDs modulate and reflect ambient LTE signals to communicate information without the need for dedicated power sources or transmitters. Cellular BackCom can achieve data rates from 1 to 10 kbps and supports communication distances of up to several hundred meters, depending on the availability and strength of the cellular signal. By leveraging existing cellular infrastructure like LTE, these systems enable low-power, long-range communication for IoT devices without the need for dedicated transmitters. Innovations such as using FSK modulation with ambient LTE signals allow devices to effectively separate backscattered signals from direct paths, enhancing reliability and performance \cite{liao2023band,elmossallamy2019noncoherent}. This approach leverages the widespread deployment of cellular networks, providing reliable communication coverage for various IoT applications, particularly suited for urban and industrial environments where robust, uninterrupted connectivity is essential, such as in smart city infrastructure, environmental monitoring, and asset tracking systems \cite{li2024intra,sheikh2021ultra}. Existing studies have focused on improving the reliability of signal backscattering over long distances, addressing challenges such as signal interference, noise, and multipath fading. Additionally, efforts have been made to optimize energy harvesting, ensuring that IoT devices can operate on harvested ambient signals alone. Extensive research has been conducted to optimize cellular backscatter systems, focusing on enhancing signal reliability, maximizing energy harvesting efficiency, and increasing data throughput \cite{ memon2019backscatter,kim2017hybrid}. Additionally, research has examined the integration of Cellular BC into 4G and 5G networks, ensuring compatibility and performance scalability in urban and industrial settings. This integration makes Cellular BackCom a critical technology for smart cities, environmental monitoring, and large-scale IoT deployments, where robust and scalable connectivity is essential \cite{ liao2023band,nawaz2021non}.

\section{Integration of 0G Networks and BackCom Technology}
This section provides a comprehensive overview of 0G networks, outlining their essential features and capabilities. It explores how integrating 0G networks with BackCom technologies addresses the critical need for ultra-low-power, low-cost connectivity in IoT applications. By combining the long-range, energy-efficient characteristics of 0G with the passive, battery-free capabilities of backscatter, this integration enables efficient data transmission in environments with limited power and infrastructure. The exploration provided aims to underscore the significance of integrating these technologies in driving sustainable, large-scale IoT deployments across sectors such as environmental monitoring, agriculture, and smart infrastructure.

\subsection{0G Networks: Overview and key features}
The exponential growth of IoT has accelerated the demand for ultra-low-power, cost-effective, and efficient connectivity solutions, renewing interest in 0G networks. Early 0G networks, developed before the advent of First-Generation (1G) cellular networks, laid the foundational principles for modern mobile communication. Although they were limited in capacity and functionality compared to their successors, 0G networks introduced key concepts that continue to influence today's technologies \cite{agrawal2022comparison}. Unlike traditional cellular networks, which prioritize high data rates and bandwidth-intensive applications, 0G networks focus on energy efficiency, extended range, and minimal infrastructure requirements. These features make them particularly well-suited for applications involving infrequent, small data transmissions, such as remote sensing, environmental monitoring, and asset tracking \cite{mahmud2019cellular}. Recently, Sigfox has redefined the term 0G to describe contemporary Low-Power Wide-Area Network (LPWAN) technologies. Modern 0G networks employ Ultra-NarrowBand (UNB) communication to provide cost-effective and energy-efficient connectivity solutions for IoT and Machine-to-Machine (M2M) communications. These networks utilize advanced modulation techniques and optimized communication protocols to enhance performance while maintaining low power consumption, significantly extending device battery life \cite{kumar2023low,moons2019efficient}.

0G networks employ optimized communication protocols that reduce both data transmission frequency and volume, significantly lowering power consumption and allowing devices to operate for extended periods without frequent maintenance or battery replacements. By leveraging low-frequency bands and UNB modulation, 0G networks achieve extensive geographical coverage, spanning several kilometers in urban areas and tens of kilometers in rural settings with minimal infrastructure investment. This long-range capability reduces the need for a dense network of BSs, lowering deployment and maintenance costs and making 0G networks ideal for applications spread over large regions. Although the data rates of 0G networks are modest, typically ranging from a few bits per second to several kilobits per second, they are well-suited for transmitting small data packets, such as sensor readings, status updates, and telemetry data. This focus on low data rates enhances energy efficiency and simplifies network architecture, making 0G networks particularly advantageous for IoT and M2M communications that require small, infrequent data transmissions \cite{jiang2024cellular}.

\begin{figure*}[h]
\centering
  \includegraphics[width=15cm]{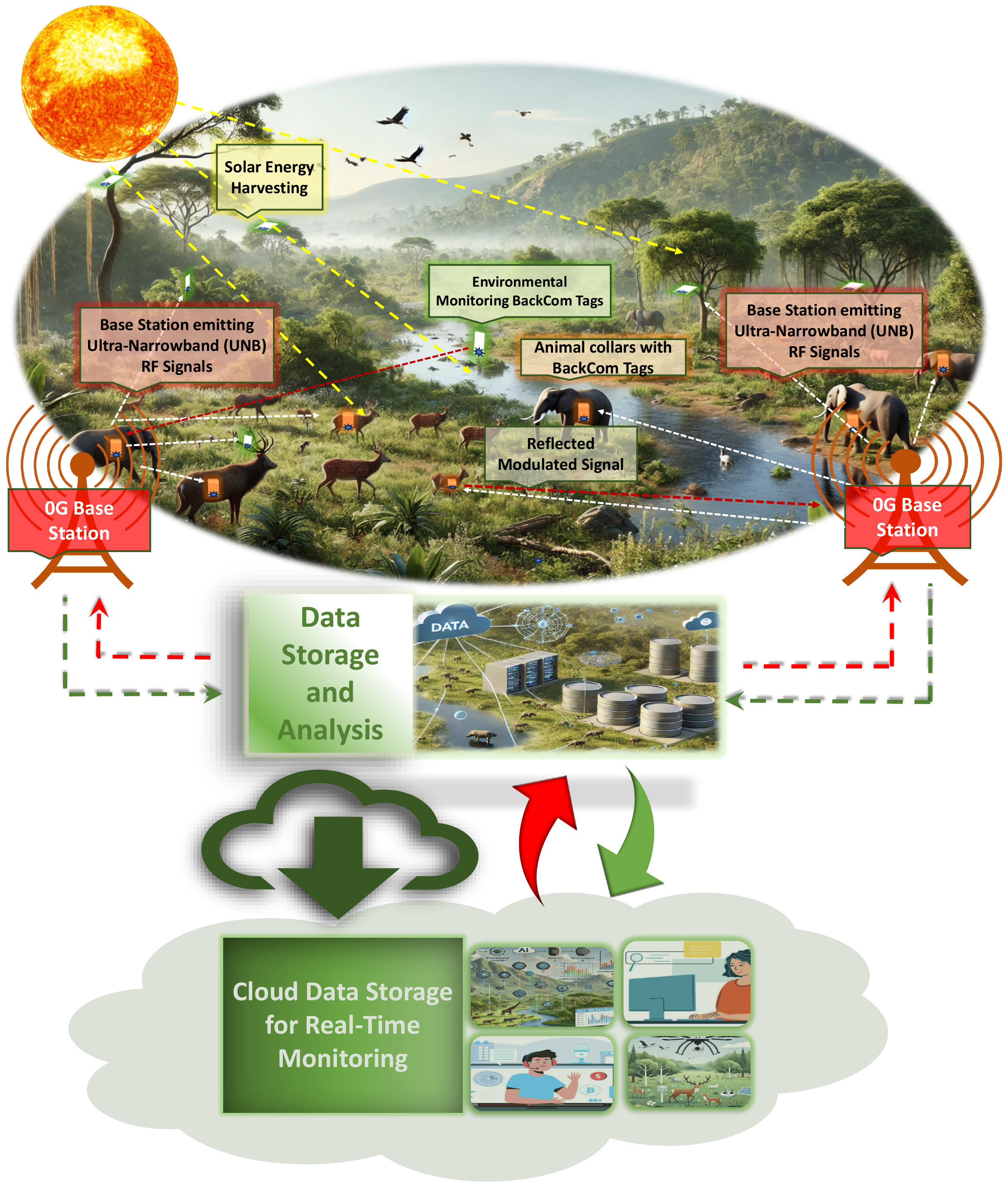}
  \caption{BackCom-Enabled 0G Network Architecture for Wildlife Monitoring.}
  \label{fig:framework}
  \end{figure*}

\subsection{Integrated Architecture of 0G network with BackCom}
The integration of 0G networks with BackCom constitutes a highly efficient and sustainable architecture tailored for extensive IoT applications. This hybrid architecture leverages the inherent strengths of both technologies to deliver long-range, low-power communication solutions with minimal infrastructure requirements to address the critical needs of modern connectivity demands. The main components and working principles are highlighted below:
\subsubsection{Primary Components}
The integrated architecture comprises several key components that collaboratively ensure seamless communication and efficient energy utilization. BSs serve as the central hubs, facilitating communication between BD (tags) and the core network. These BS units are equipped with UNB transceivers capable of generating and transmitting Continuous Wave (CW) RF signals over extensive geographical areas \cite{naik2018lpwan}. This broad coverage minimizes the need for a dense network of BSs, thus reducing both the deployment and maintenance costs.

BDs (Tags) are the nearly passive sensors or actuators within this architecture. Unlike traditional devices that generate their own RF signals, BDs transmit data by reflecting and modulating the incoming RF signals from the BS. This method drastically reduces energy consumption, allowing these devices to operate for extended periods on minimal power sources such as small batteries or energy harvested from the environment (e.g., solar or thermal energy). The core network is responsible for managing data processing, storage, and integration with external systems or applications. It interfaces with cloud platforms, databases, and user applications, ensuring that the data aggregated from multiple BDs is processed efficiently and securely. Antennas play a crucial role in facilitating the transmission and reception of RF signals between the BS and BDs. High-gain antennas at the BS maximize coverage and signal strength, while compact and efficient antennas on BDs optimize their reflection and modulation capabilities \cite{mushtaq2020design}. Furthermore, communication protocols define the rules and standards for data transmission, modulation, and error handling within the network. These protocols are optimized for low data rates and minimal energy consumption, ensuring reliable data transmission despite the inherent low power and potential interference challenges \cite{chaudhari2020lpwan}.

% Please add the following required packages to your document preamble:
% \usepackage{graphicx}
\begin{table*}[h]
\caption{Comparison of BackCom-Enabled 0G Network and Traditional BackCom Systems}
\label{tab:comparison}
\renewcommand{\arraystretch}{1.2} % Adjust row spacing
\resizebox{\textwidth}{!}{%
\begin{tabular}{|l|l|l|}
\hline
\textbf{Features} &
  \textbf{BackCom-Enabled 0G Networks} &
  \textbf{Traditional BackCom Systems} \\ \hline
\textbf{Power Efficiency} &
  \begin{tabular}[c]{@{}l@{}}Ultra-low power due to integration \\ with 0G’s UNB technology\end{tabular} &
  \begin{tabular}[c]{@{}l@{}}Low power but depends entirely on ambient RF \\ sources or a dedicated RF reader\end{tabular} \\ \hline
\textbf{Data Transmission Range} &
  \begin{tabular}[c]{@{}l@{}}Extended range supported by 0G \\ infrastructure and ambient signals\end{tabular} &
  \begin{tabular}[c]{@{}l@{}}Limited range, typically short-distance due to \\ reliance on local ambient RF sources\end{tabular} \\ \hline
\textbf{Infrastructure Requirements} &
  \begin{tabular}[c]{@{}l@{}}Leverages 0G base stations, allowing \\ wider deployment across large areas\end{tabular} &
  \begin{tabular}[c]{@{}l@{}}Requires dedicated RF readers or ambient \\ RF sources such as Wi-Fi or TV signals\end{tabular} \\ \hline
\textbf{Scalability} &
  \begin{tabular}[c]{@{}l@{}}Highly scalable, can support large \\ sensor networks in remote areas\end{tabular} &
  \begin{tabular}[c]{@{}l@{}}Limited scalability as multiple readers may \\ be required for large coverage\end{tabular} \\ \hline
\textbf{Reliability of Communication} &
  \begin{tabular}[c]{@{}l@{}}More reliable, supported by 0G \\ network’s consistent signal coverage\end{tabular} &
  \begin{tabular}[c]{@{}l@{}}Less reliable in remote areas without ambient \\ RF sources\end{tabular} \\ \hline
\textbf{Deployment Cost} &
  \begin{tabular}[c]{@{}l@{}}Cost-effective for large-scale deployments, \\ minimal infrastructure with 0G support\end{tabular} &
  \begin{tabular}[c]{@{}l@{}}Low-cost but may require additional RF sources \\ or infrastructure for large areas\end{tabular} \\ \hline
\textbf{Data Security} &
  \begin{tabular}[c]{@{}l@{}}Enhanced security through centralized \\ 0G data handling and encryption\end{tabular} &
  \begin{tabular}[c]{@{}l@{}}Limited security, data is often processed locally \\ and may be less secure\end{tabular} \\ \hline
\textbf{Real-Time Monitoring} &
  \begin{tabular}[c]{@{}l@{}}Supports real-time monitoring with minimal \\ power use\end{tabular} &
  \begin{tabular}[c]{@{}l@{}}Limited real-time capability due to range and \\ dependency on ambient sources\end{tabular} \\ \hline
\textbf{Use Case Suitability} &
  \begin{tabular}[c]{@{}l@{}}Ideal for remote wildlife monitoring, \\ large-scale environmental deployments\end{tabular} &
  \begin{tabular}[c]{@{}l@{}}Suitable for smaller, controlled environments \\ with access to infrastructure\end{tabular} \\ \hline
\end{tabular}%
}
\end{table*}

\subsubsection{Working Principles}
Several fundamental principles underpin the operational efficiency of this integrated architecture. Initially, the BS generates a CW-RF signal using UNB modulation, which is then transmitted over a wide area. Upon receiving this signal, the BD activates its modulation circuitry to reflect and modulate the incoming signal, encoding data by altering its reflection properties (e.g., changing antenna impedance). This modulated signal is then reflected back to the BS, where advanced signal processing techniques demodulate and decode the transmitted data despite the low signal strength and potential noise interference. Once demodulated, the data is forwarded from the BS to the core network for processing and integration with other systems, making it accessible for various applications such as environmental monitoring, asset tracking, or smart agriculture. The integration of BC enhances energy efficiency by reducing the need for active signal generation and extending connected devices' battery life, thereby promoting sustainability and reducing maintenance costs.

\subsection{Waste Factor (WF) Analysis of BackCom-enabled 0G system}
The concept of WF has recently emerged as a novel metric for evaluating the power efficiency of communication systems, particularly in resource-constrained IoT applications \cite{10604822}. It quantifies the proportion of energy wasted relative to useful data transmission, with a lower WF indicating more efficient energy use. This metric is especially relevant in BackCom-enabled 0G networks, which prioritize ultra-low power communication by utilizing ambient RF signals instead of actively generating transmissions. By integrating energy-harvesting capabilities with optimized protocols, these systems significantly reduce energy consumption and rely on reflected signals for data transmission without dedicated power sources. In a BackCom-enabled 0G system, achieving power efficiency is critical due to the passive or semi-passive nature of BDs. The WF, as defined in Equation \ref{eq:first}, quantitatively illustrates how the integrated system minimizes energy consumption:

\begin{equation}
\label{eq:first}
\mathrm{WF} = \frac{P_{\text{total}} - P_{\text{useful}}}{P_{\text{useful}}} = \frac{P_{\text{wasted}}}{P_{\text{useful}}}
\end{equation}

In the context of a BackCom-enabled 0G system, we consider both the power consumed by the BS and the BDs. However, since BDs do not actively generate RF signals but instead modulate and reflect existing ones, their power consumption remains significantly lower than that of conventional devices.

\textbf{System Power Components Definition:}
The parameters used for BackCom-Enabled 0G Systems are defined as follows:
\begin{itemize}
    \item $P_{\text{BS}}$: Power consumed by the base station to transmit the CW signal.
    \item $P_{\text{BD}_i}$: Power consumed by the $i$-th BD for modulation and minimal processing.
    \item $P_{\text{total}}$: Total power consumed by the system, calculated as shown in Equation \ref{Eq:second}:
    \begin{equation}
    \label{Eq:second}
    P_{\text{total}} = P_{\text{BS}} + \sum_{i=1}^{N} P_{\text{BD}_i}
    \end{equation}
    \item $P_{\text{useful}}$: Total power effectively used for transmitting useful data, defined as the sum of the power of the backscattered signals from all BDs, given in Equation \ref{Eq:Useful}:
    \begin{equation}
    \label{Eq:Useful}
    P_{\text{useful}} = \sum_{i=1}^{N} P_{\text{backscatter}_i}
    \end{equation}
    \item $N$: Total number of BDs in the network.
\end{itemize}

Thus, the WF for the BackCom-Enabled 0G system can be expressed as shown in Equation \ref{Eq:OGWF}:

\begin{equation}
\label{Eq:OGWF}
\mathrm{WF}_{\text{0G}}^{\text{BackCom}} = \frac{P_{\text{BS}} + \sum_{i=1}^{N} P_{\text{BD}_i} - \sum_{i=1}^{N} P_{\text{backscatter}_i}}{\sum_{i=1}^{N} P_{\text{backscatter}_i}}
\end{equation}

Given that $P_{\text{BD}_i}$ is minimal and $P_{\text{backscatter}_i}$ depends on the modulated reflection of the BS’s signal, the WF can be significantly reduced by optimizing $P_{\text{BS}}$ and maximizing $P_{\text{backscatter}_i}$. Accordingly, the system efficiency is defined as Equation \ref{Eq:SE}:

\begin{equation}
\label{Eq:SE}
\eta_{\text{system}} = \frac{P_{\text{useful}}}{P_{\text{total}}} = \frac{\sum_{i=1}^{N} P_{\text{backscatter}_i}}{P_{\text{BS}} + \sum_{i=1}^{N} P_{\text{BD}_i}}
\end{equation}

Thus, the Waste Factor can also be expressed in terms of system efficiency:

\begin{equation}
\mathrm{WF}_{\text{0G}} = \frac{1}{\eta_{\text{system}}} - 1
\label{eq:efficiencyWF}
\end{equation}

Accordingly, it shows that as system efficiency approaches 100\%, the WF approaches zero, indicating minimal energy waste. Optimizing BackCom-enabled 0G systems for greater energy efficiency is crucial, as reducing power consumption and leveraging backscatter to utilize ambient RF signals can significantly lower energy usage. This makes these systems ideal for large-scale IoT deployments focused on ultra-low power communication. As IoT networks expand, achieving near-zero WF through energy-harvesting and backscatter techniques will be vital for sustaining scalable and environmentally friendly systems.

\subsection{Case Scenario: Wildlife Monitoring with 0G-BackCom Integration}
To further illustrate the integrated architecture of the BackCom-enabled 0G system, Figure \ref{fig:framework} depicts a wildlife monitoring architecture. The 0G network is combined with BackCom technology to facilitate an efficient, large-scale environmental monitoring system in wildlife reserves or natural habitats. In this scenario, a vast area is equipped with multiple animal tracking and environmental monitoring sensors to observe animal health, movement, and ecological conditions. These sensors, implemented as BDs, are strategically placed in animal collars or environmental nodes to ensure wide-area coverage.

The BDs are designed to operate efficiently and with minimal energy consumption, taking full advantage of the 0G network's ultra-low power communication to ensure long-term operation in remote wildlife habitats with minimal maintenance. To achieve this, each BD is equipped with compact, efficient antennas that are optimized for reflecting and modulating incoming RF signals from the BSs, enabling the measurement of specific environmental parameters such as temperature, humidity, and animal movement. The BSs are installed at critical points around the wildlife reserve, providing extensive coverage with minimal infrastructure. These BSs continuously emit UNB RF signals, which the BDs receive. Upon receiving the CW RF signal from a BS, each BD encodes data by altering its reflection properties, allowing it to transmit data back to the BS without generating its own RF signal. This reflection-based method minimizes energy consumption, allowing BDs to function for years on minimal power sources, such as small batteries or solar energy harvested from the environment. The BSs decode these reflected signals and forward the data to a core network, where it is processed and integrated with cloud-based management systems. Wildlife researchers and conservationists can then access real-time data through control panels, enabling informed decisions on animal health, migration patterns, and habitat conditions while minimizing maintenance efforts and maximizing operational efficiency.

Consequently, the inherent advantages of 0G networks can benefit significantly from the integration of BackCom technology, further amplifying their potential to improve the performance of ultra-low power applications. Integration of BackCom technology with the optimized communication protocols of 0G networks and UNB allows for the creation of nearly passive devices that require minimal power for operation. This results in extended battery life for tracking devices, ensuring several years of operation without the need for replacements, which is essential for wildlife tracking in remote or difficult-to-reach areas. The low-power nature of BackCom also complements 0G's wide geographical coverage, enabling the deployment of vast sensor networks across large wildlife reserves, national parks, or other protected areas without significant infrastructure investments. The combination of 0G networks and BackCom simplifies network architecture, enhances scalability, and reduces maintenance costs, making it an ideal solution for large-scale wildlife monitoring, habitat conservation, and biodiversity research, where longevity and reliability are paramount \cite{wu2022survey,kellogg2016passive}. Table \ref{tab:comparison} provides a detailed comparison of the BackCom-Enabled 0G network and traditional BackCom systems to highlight the advantages of integrating it with the 0G network.

% \section{Waste Factor (WF) in 0G-BackCom Systems}
% This section presents a comprehensive methodology for calculating the WF and demonstrates its application in assessing the efficiency of the communication system. By providing a detailed, step-by-step breakdown of the WF metric, we illustrate how it serves as a powerful quantitative tool to measure energy utilization within integrated frameworks, particularly in BackCom-enabled 0G networks where ultra-low power consumption is critical.

% \subsection{Overview and Significance of Waste Factor (WF) for BackCom-Enabled 0G Networks}

\section{Potential Applications}
This section explores the potential applications of Backscatter-Enabled 0G Networks, highlighting their role in transforming wireless communication across various domains. It examines how these networks can be employed to support ultra-low-power green IoT architectures, enabling scalable and sustainable solutions for smart cities, environmental monitoring, healthcare, and industrial automation.

\begin{figure*}
\centering
  \includegraphics[width=18cm]{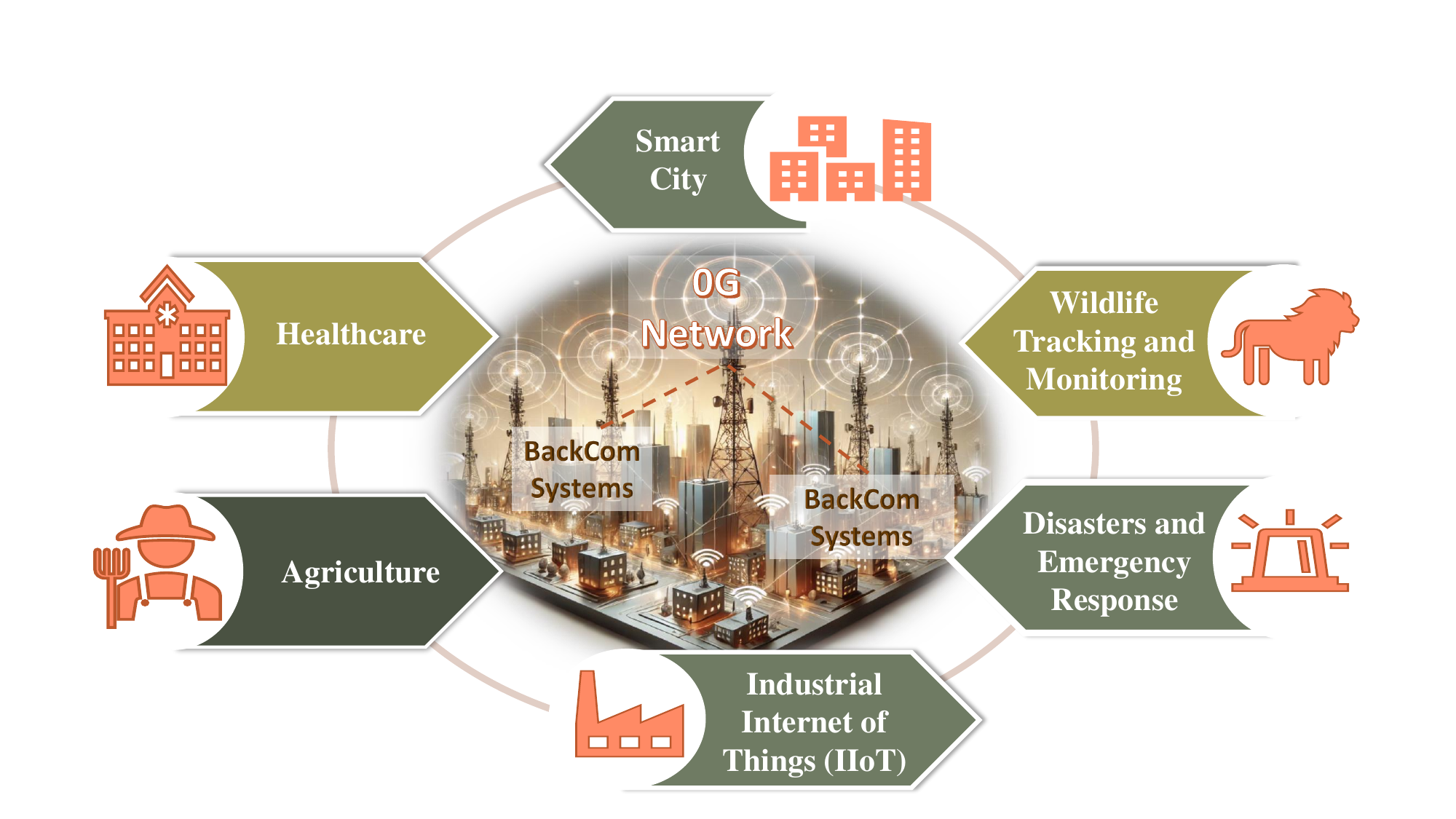}
  \caption{Applications of Backscatter-Enabled 0G Networks.}
  \label{fig:applications}
  \end{figure*}

\subsection{Healthcare Applications}
BackCom technology-enabled 0G networks offer a promising solution to improve healthcare by utilizing the extended range and low power features of 0G networks along with the passive and ultralow power communication strengths of BackCom. This integration can play a significant role in remote patient monitoring for managing chronic diseases, post-operative care, and elderly care by facilitating the deployment of a vast array of low-power sensors and wearable devices that monitor vital signs such as heart rate, blood pressure, glucose levels, and oxygen saturation in real-time \cite{jameel2019applications}. The ultra-low power consumption allows these devices to operate for years without the need for battery replacement, improving patient comfort and reducing maintenance costs. 0G networks provide robust connectivity over large areas, ensuring uninterrupted data transmission even in remote or underserved locations. The passive nature of BDs allows for the deployment of numerous sensors without significant infrastructure investments. This is particularly advantageous for smart medical devices, such as smartwatches, fitness trackers, and implantable sensors, which can leverage BackCom to offer continuous health monitoring without frequent charging, leading to extended device usability and adherence to monitoring protocols \cite{zhang2016enabling}. In addition, passive communication modules help to miniaturize wearable devices, improving both patient compliance and comfort. Integration with 0G networks ensures that health data is reliably transmitted to healthcare providers, enabling timely interventions \cite{galappaththige2022link,ren2023toward}.

Beyond patient care, hospitals and healthcare facilities can optimize asset tracking and inventory management by using Backscatter-enabled RFID tags integrated with 0G networks. This provides precise location data and status updates on medical equipment, reducing the risk of loss or theft while optimizing stock levels to ensure critical supplies are always available. Additionally, the broad coverage of 0G networks requires fewer base stations, thereby reducing deployment and maintenance costs. This integration also facilitates early detection of environmental anomalies, preventing equipment failures, and ensuring compliance with healthcare standards \cite{yao2020backscatter,ahmed2024noma}.

\subsection{Industrial Internet of Things (IIoT)}
The Industrial Internet of Things (IIoT) applications require robust, scalable, and energy-efficient communication solutions to facilitate advanced manufacturing, predictive maintenance, and optimized operational efficiencies in industrial environments. Integrating 0G networks with BackCom leverages the extended range and low-power characteristics of 0G with the ultra-low-power, passive communication capabilities of backscatter, resulting in a highly efficient and scalable communication infrastructure. In industrial settings, numerous sensors and actuators are deployed across extensive areas to monitor parameters such as temperature, vibration, pressure, and machine status. This integration enables these devices to operate with minimal power consumption by using ambient RF signals or dedicated RF sources that enable BDs to transmit data without the need for active RF generation, significantly extending battery life or enabling energy-harvesting solutions. This extended operational lifespan reduces maintenance costs and improves the sustainability of IIoT deployments \cite{niu2019overview,zargari2023signal}.

Furthermore, 0G networks use low-frequency bands and UNB modulation techniques to provide the range and penetration capabilities necessary for reliable connectivity across large industrial sites. Integrating BackCom with existing 0G infrastructure enhances overall network robustness. This ensures that data from remote or hard-to-reach sensors is reliably transmitted to central monitoring systems without requiring a dense network infrastructure. In addition, implementing extensive network infrastructure in industrial settings can be costly, however, integrating 0G networks with BackCom reduces infrastructure expenses by decreasing the need for BSs and utilizing existing RF sources for data transmission. This cost-effective approach enables the scalable deployment of large sensor networks, allowing comprehensive monitoring and control across diverse industrial processes without incurring significant capital expenditures \cite{feltrin2019narrowband,rastogi2020narrowband}.

\subsection{Agriculture Applications}
The integration of 0G networks with BackCom technology enables the large-scale deployment of sensors across farmlands, providing real-time monitoring of critical factors such as soil moisture, nutrient levels, temperature, and humidity. These ultra-low-power sensors operate passively, eliminating the need for frequent battery replacements. By utilizing 0G networks to transmit data over extensive areas, farmers gain continuous insights into crop health and soil conditions, enabling them to make data-driven decisions on irrigation, fertilization, and pest control \cite{jiang2023backscatter}. Traditional network infrastructure can be prohibitively expensive in rural and expansive agricultural regions. However, the combination of 0G networks and BackCom significantly reduces the need for extensive infrastructure, cutting deployment and maintenance costs. The passive nature of BDs allows them to leverage existing RF signals for data transmission, further lowering costs while maintaining reliable connectivity \cite{daskalakis2018uw,daskalakis2017ambient}.

In addition to crop management, 0G and BackCom technology can also enhance livestock and equipment tracking. By utilizing backscatter-enabled RFID tags for animals, farmers can monitor their location, health, and behavior patterns in real-time \cite{henry2018automated}. Similarly, agricultural machinery can be tracked to optimize usage, minimize downtime, and improve maintenance schedules, thereby enhancing overall operational efficiency. The ultra-low power consumption of BackCom technology also supports the deployment of extensive sensor networks without significantly increasing energy demand. This allows farmers to monitor environmental conditions and implement precision agriculture techniques that conserve water, reduce fertilizer usage, and minimize pesticide application, ultimately promoting eco-friendly farming practices. Furthermore, this integrated network facilitates the early detection of pests and diseases by continuously monitoring microclimatic conditions and plant health indicators. Real-time data collection can trigger automated alerts, enabling farmers to take preventive actions before problems escalate, thus reducing crop loss and boosting yields. Moreover, deploying Backscatter-based soil moisture sensors connected through 0G networks allows farmers to optimize irrigation schedules and conserve water resources to enhance soil quality.

\subsection{Smart City Applications}
The integration of 0G networks with BackCom technology has the potential to revolutionize smart cities by providing a cost-effective, scalable, and energy-efficient communication system. This combination can significantly enhance urban management, optimize resource utilization, and promote sustainable development. Ultra-low-power sensors can be deployed to monitor water distribution, gas pipelines, and electricity grids in real-time, allowing cities to detect leaks and faults efficiently, thus streamlining resource management and reducing maintenance needs. In addition, backscatter-enabled sensors in waste bins can optimize waste collection by monitoring fill levels, reducing fuel consumption, and lowering operational costs \cite{balid2017development}.

Environmental monitoring can also benefit significantly from this integration, as passive sensors continuously track air quality, temperature, and noise levels, enabling city planners to implement strategies for reducing pollution and enhancing urban living conditions. In transportation, real-time data collected from passive RFID tags on vehicles and infrastructure can optimize traffic flow and parking availability, reducing congestion and emissions. Moreover, smart city lighting systems can adjust streetlights based on ambient conditions and pedestrian movement, thereby reducing electricity consumption and operational expenses. Moreover, in building management, it has the potential to support real-time monitoring of energy usage, temperature control, and occupancy, optimizing efficiency and reducing operational costs. This combination of 0G networks and BackCom technology creates a robust foundation for more innovative, safer, and more sustainable urban environments \cite{bletsas2018art,yamashita2019ultra}.

\subsection{Wildlife Tracking and Monitoring}
Traditional tracking systems often require frequent battery changes or a constant power supply, which can be challenging to maintain in remote or inaccessible wildlife habitats. By integrating BackCom technology with 0G networks, wildlife trackers can send data over long distances with minimal power consumption. These networks enable the transmission of small amounts of data, such as location, temperature, and environmental conditions, without the need for a dense communication infrastructure. Passive RFID tags or backscatter-enabled sensors can be attached to animals to track their movements. These tags send back data based on ambient RF signals reflected off the environment, allowing the animals' positions to be mapped as they move through different areas. This information is crucial for understanding migration patterns, territorial behavior, and animal movement within protected areas or across borders \cite{ross2022wildtrack}. In addition to tracking movement, backscatter-enabled sensors can monitor other environmental conditions, including temperature, humidity, and air quality around the animal. This data helps researchers understand how wildlife is adapting to changes in their environment, including the effects of climate change or human activity. For example, monitoring the behavior of certain species (e.g., nesting, feeding, or mating) can provide insights into how they are coping with habitat loss or the presence of humans \cite{pereira2023rfid}.

Unlike traditional GPS or satellite tracking systems, the passive communication capabilities of BackCom technology do not rely on active signal emissions from the animal's tracker. This makes it less intrusive and reduces the likelihood of disturbing the animals, which is essential in wildlife conservation, particularly for sensitive or endangered species. Moreover, it supports real-time communication from wildlife sensors to central monitoring systems. For instance, if an animal crosses a boundary, enters a dangerous area, or is in distress, the system can immediately alert park rangers or wildlife managers. Access to real-time data helps authorities respond quickly to potential poaching incidents, human-wildlife conflicts, or environmental hazards like fires or floods. This integrated system approach is also cost-effective, as deploying traditional tracking systems is generally expensive, particularly in vast or difficult-to-access areas. However, BackCom-enabled 0G networks reduce the need for extensive infrastructure, allowing for scalable deployments that cover large areas with fewer resources, making wildlife monitoring more affordable and sustainable in the long term \cite{galappaththige2022link,baratchi2013sensing}.

\subsection{Disasters and Emergency Response}
BackCom-enabled 0G networks can significantly enhance early warning systems for natural disasters such as earthquakes, floods, and wildfires by deploying passive sensors in vulnerable areas. These sensors can continuously monitor environmental indicators such as soil moisture, river levels, and seismic activity, transmitting real-time data over large areas. This allows authorities to issue early alerts and evacuate at-risk populations, potentially saving lives. In addition, after a disaster, a key priority is assessing damage to critical infrastructure such as bridges, buildings, and roads. By using this integrated network, authorities can monitor structural integrity in real-time. BackCom enables sensors to detect unusual vibrations, cracks, or shifts, helping to quickly identify areas at risk of collapse and enabling faster data-driven decisions on which structures are safe and which require immediate attention \cite{badirkhanli2020rescue}.

Since 0G networks offer robust, long-range connectivity that does not rely heavily on physical infrastructure, they are ideal for communication in remote or heavily damaged regions. Passive sensors and low-power devices ensure that emergency communication remains functional, even in areas with limited electricity access. Efficient disaster management also requires the timely allocation of resources such as food, medical supplies, and personnel. Backscatter-enabled RFID tags can track inventory in real-time, ensuring that relief supplies reach suitable locations swiftly. Additionally, these tags can monitor the movement of emergency vehicles and personnel, optimizing logistics and reducing response times. Moreover, in search and rescue operations, 0G and BackCom technology are invaluable. Passive sensors deployed in disaster zones can detect signs of life, such as movement, body heat, or sounds, and communicate over long distances with minimal infrastructure. This helps rescuers locate trapped individuals quickly, even in environments where GPS and cellular networks may not function reliably \cite{wang2022resource}. In addition, post-disaster scenarios often bring secondary hazards, including landslides, gas leaks, or contaminated water supplies. Backscatter sensors can continuously monitor environmental conditions to detect such hazards early, enabling authorities to take preventive measures. For example, sensors can track air quality to detect harmful gases or monitor water sources for contamination, ensuring public safety in the aftermath of a disaster \cite{zhou2024enhanced}.

\section{Open Challenges and Future Research Directions}
This section provides an in-depth analysis of the open challenges and future research directions for Backscatter-Enabled 0G Networks, focusing on key areas that require further exploration to realize their full potential.

\subsection{Energy Harvesting and Sustainability}
While BackCom reduces device energy consumption by eliminating active signal generation, ensuring a sustainable power supply for BDs in 0G networks remains challenging, especially in remote areas with limited energy harvesting options. These devices often rely on small batteries or ambient energy, but the inconsistency of sources such as light, thermal gradients, and mechanical vibrations can disrupt continuous operation. Despite their low energy needs, BDs still require a stable power supply for long-term functionality without frequent maintenance. Achieving sustainability and reliability in such conditions calls for innovative approaches in energy harvesting and power management tailored to BackCom-enabled 0G networks \cite{ji2020joint,tang2021self}.

Consequently, to address these challenges, future research should prioritize identifying alternative energy sources beyond conventional solar power, such as ambient light, thermal gradients, and mechanical vibrations, with thermoelectric and piezoelectric generators efficiently converting these into usable energy to improve system sustainability. Moreover, advancing material science, such as developing high-efficiency photovoltaic cells and effective piezoelectric materials, is essential to maximize energy conversion \cite{he2024bulk,ma2019sensing}. Future work should also optimize communication protocols specifically for low-power BackCom systems that can reduce energy consumption by minimizing transmission frequency, using efficient modulation schemes, and applying data aggregation. Lightweight data compression algorithms that align with the limited computational capabilities of BDs can further lower energy costs. Adaptive transmission mechanisms adjusting parameters (e.g., power levels and data rates based) on real-time energy availability can optimize network efficiency, with ML models predicting optimal strategies. Future research should also focus on fully battery-free BDs that rely exclusively on harvested energy, necessitating ultra-low-power hardware design to sustain operations without batteries. Integrating passive RFID technology with BackCom offers a viable approach, as these systems harness energy from reader signals, establishing the foundation for sustainable battery-free device designs within 0G networks \cite{xu2018practical,jiang2023backscatter,jameel2020noma}.

\subsection{Advanced Modulation and Signal Processing Techniques}
The integration of advanced modulation and signal processing techniques within BackCom-enabled 0G networks presents considerable technical challenges, primarily due to the stringent requirements of low-power, long-range communication. Traditional modulation schemes, which typically demand significant power for signal generation and transmission, are incompatible with BackCom systems, where devices rely on reflecting and modulating existing RF signals rather than generating their own. Consequently, developing ultra-low-power modulation techniques capable of encoding data with minimal energy input while ensuring reliable communication over extensive distances is essential \cite{yang2017modulation}. Additionally, BackCom systems are characterized by weak reflected signals that are highly susceptible to interference from environmental factors and other wireless technologies operating in shared spectrum, resulting in a low signal-to-noise ratio. This complicates the demodulation and decoding processes, necessitating sophisticated signal-processing algorithms for accurate data interpretation. Moreover, environmental variability, including multipath fading and shadowing, further compromises consistent signal quality, thereby challenging the overall reliability of 0G-BackCom networks. Balancing the need for high data integrity and robust communication with BackCom’s stringent energy constraints remains a critical barrier to advancing 0G network architectures \cite{xu2018practical,ma2018blind}.

To overcome these challenges, future research should focus on developing innovative modulation schemes and enhancing signal processing algorithms tailored to the specific requirements of these systems. Novel modulation techniques, such as differential modulation and spread spectrum methods, should be explored to improve resilience against interference and multipath effects while maintaining low power consumption. Additionally, adaptive modulation strategies that dynamically adjust parameters based on real-time channel conditions and energy availability can significantly enhance communication reliability and efficiency. Advanced demodulation algorithms leveraging ML and Artificial Intelligence (AI) should be developed to more accurately decode weak and noisy backscattered signals, thereby improving data integrity and reducing error rates. Effective interference mitigation is also crucial, necessitating the creation of algorithms that can distinguish and filter unwanted signals from desired backscatter transmissions \cite{rezaei2023coding}. Furthermore, energy-efficient signal processing frameworks must be designed to minimize computational overhead, ensuring that processing tasks do not negate the energy savings achieved through BackCom. The integration of edge computing capabilities can also play a pivotal role by offloading complex signal processing tasks from low-power BDs to more capable edge nodes, thereby enhancing overall network performance without compromising energy efficiency \cite{asif2022energy,bletsas2010improving}.

\subsection{Efficient Spectrum Utilization and Network Coordination}
BDs reflect and modulate existing RF signals rather than generating their own, leading to shared spectrum usage between active and passive devices. This shared usage can result in substantial interference and congestion, particularly in densely deployed environments where numerous BDs operate simultaneously within the same frequency bands \cite{al2021performance}. The use of UNB modulation, while beneficial for long-range communication and low power consumption, occupies specific frequency segments, making spectrum management crucial to avoid overlap and interference. The decentralized nature of many 0G network deployments complicates network coordination, as there is often no centralized control mechanism to manage spectrum access among the numerous passive devices. Efficient spectrum utilization is essential to maximize network capacity and reliability while minimizing energy consumption and operational costs. The dynamic and unpredictable nature of IoT applications, where devices may frequently join or leave the network, exacerbates these challenges, necessitating robust and adaptable spectrum management and coordination solutions \cite{liang2022backscatter}.

Consequently, these challenges necessitate that future research focus on developing advanced spectrum management and coordination strategies personalized to the specific needs of these systems. Dynamic spectrum allocation should be explored to adaptively assign frequency resources based on real-time network demand and interference levels to avoid congestion and optimize spectrum usage by ensuring that BDs operate within the least crowded frequency bands. Additionally, research into cognitive radio systems can enable BackCom devices to intelligently detect and utilize available spectrum segments without causing interference to other devices \cite{guo2019cognitive}. Future studies should also pay attention to the development of coordinated multi-device communication protocols to facilitate orderly access to the shared spectrum, preventing data collisions, and optimizing data flow among numerous passive devices. Decentralized coordination mechanisms, such as those leveraging distributed algorithms or blockchain technology, can offer scalable and resilient solutions by enabling seamless network coordination without relying on a single point of control \cite{liang2020symbiotic}. Integrating ML-based spectrum management can further enhance these systems by predicting spectrum usage patterns and dynamically optimizing allocation based on changing network conditions and device behaviors. In addition, standardization efforts are crucial to establish protocols and guidelines for spectrum usage and device coordination in BackCom-enabled 0G networks, promoting interoperability and facilitating the widespread adoption of robust, scalable, and energy-efficient green IoT applications \cite{xu2023state,khan2021learning}.

\subsection{Latency Optimization and Quality of Service (QoS) Management}
The passive transmission mode inherent in BackCom-enabled 0G networks introduces variability in communication latency, making it challenging to achieve consistent and predictable data delivery. The ultra-low power consumption and limited computational resources of BDs also restrict the deployment of sophisticated QoS management protocols typically employed in more active communication systems. This poses significant difficulties in ensuring reliable and timely data transmission, especially for real-time applications such as industrial automation and emergency response systems. The shared spectrum usage and potential interference from multiple BDs further intensify latency issues as signal collisions and retransmissions increase delays and degrade overall network performance \cite{jameel2020low,liu2019next}. Achieving low-latency and high-QoS communication while maintaining the energy-efficient and long-range objectives of 0G networks necessitates innovative solutions targeting both the physical and network layers.
Future research must focus on developing adaptive QoS protocols capable of dynamically adjusting parameters based on real-time network conditions and application requirements. These protocols should optimize bandwidth allocation, prioritize time-sensitive data packets, and employ traffic-shaping techniques to minimize latency. Moreover, designing latency-aware routing algorithms is essential to ensure data packets are transmitted via the most efficient paths, reducing transmission delays and avoiding network congestion. The integration of edge computing can further reduce latency by processing data closer to the source, thus decreasing transmission distances and accelerating response times. Additionally, leveraging ML techniques to predict and mitigate latency by analyzing traffic patterns and network behaviors can enable proactive adjustments to network configurations. Research into energy-efficient QoS mechanisms is also critical to balance the stringent energy limitations of BDs with the need for robust QoS guarantees, ensuring efficient and reliable communication in 0G network deployments \cite{xu2020energy,hassan2020statistical}.

\subsection{Emerging Technology Integration for Intelligent BackCom-Enabled 0G Networks}
Integrating emerging technologies into BackCom-enabled 0G networks presents several critical challenges that must be addressed to improve system intelligence, efficiency, and security. One of the primary challenges is achieving interoperability, as integrating technologies such as AI, edge computing, and blockchain with BackCom necessitates seamless communication across diverse platforms and protocols. Ensuring that these technologies can function cohesively without adding significant overhead or complexity is essential for system reliability \cite{nie2024qos}. However, the ultra-low-power and limited computational capabilities of BackCom devices present a barrier to incorporating resource-intensive AI and ML algorithms. Implementing advanced data processing and decision-making capabilities on such constrained devices while maintaining their energy efficiency remains a difficult task. Security and privacy concerns are also heightened with the integration of these technologies, as they may introduce vulnerabilities or require additional power for cryptographic operations. Moreover, scalability is a pressing issue, with the need to manage an increasing number of interconnected devices while supporting real-time data processing and analytics \cite{han2017wirelessly,gong2020deep,xu2023state}.

Future research should focus on developing lightweight AI and ML algorithms optimized for low-power BackCom devices. This includes creating efficient neural network architectures and employing techniques such as model compression and federated learning to enable on-device intelligence without excessive energy consumption. Future research into edge computing integration is also crucial, exploring decentralized processing frameworks that can offload computationally intensive tasks from BackCom devices to nearby edge nodes, thereby reducing latency and conserving device energy \cite{du2023computation}. Additionally, secure blockchain implementations should be investigated to provide robust data integrity and authentication mechanisms without imposing heavy energy costs, which could involve designing lightweight consensus algorithms tailored for resource-constrained environments . Establishing interoperability standards and protocols is vital to ensuring seamless communication between heterogeneous technologies within the BackCom-enabled 0G ecosystem. Moreover, scalable data management and analytics solutions are required to efficiently process the increasing volume of data generated by these networks, enabling real-time insights and decision-making \cite{luo2024symbiotic,luo2024convergence}.

% \section{Standardization Progress}
% Both 3GPP and IEEE are currently considering standardization for ambient BC. Their classical architectures and topologies are illustrated in Fig. .

\section{Conclusion}
The presented study explores the transformative potential of integrating BackCom technologies with 0G networks to advance the Green IoT systems. By leveraging the ultra-low-power capabilities of BackCom, coupled with the scalability and ubiquity of 0G networks, this integration offers a sustainable framework for addressing the energy and connectivity demands of next-generation IoT ecosystems. The comprehensive analysis presented in this paper highlights the architectural and operational synergies between BackCom and 0G networks, emphasizing their ability to enhance performance, minimize energy waste, and support massive connectivity with minimal ecological impact. The introduction of the WF metric provides a holistic approach to evaluating energy efficiency, enabling a more nuanced understanding of resource utilization and guiding the design of greener communication systems. The exploration of the applications of BackCom-enabled 0G networks showcases their potential to address real-world challenges. These applications demonstrate the versatility and transformative capability of this integrated framework in fostering resilient and environmentally sustainable smart systems. Despite these advancements, challenges remain in areas such as energy harvesting, interference management, spectrum utilization, and seamless integration with emerging technologies. Addressing these open issues is critical to unlocking the full potential of BackCom-enabled 0G networks. To this end, we have outlined future research directions to ensure the robustness, scalability, and sustainability of these systems. This study provides a roadmap for leveraging BackCom and 0G networks to build environmentally sustainable and scalable IoT ecosystems, paving the way for innovative solutions that balance technological progress with ecological responsibility.
% \begin{thebibliography}{1}
% \bibitem{ref1} X. Liu and N. Ansari, "Toward green IoT: Energy solutions and key challenges," \textit{IEEE Commun. Mag.}, vol. 57, no. 3, pp. 104–110, Mar. 2019.
% \bibitem{ref2} C. Xu, L. Yang, and P. Zhang, "Practical backscatter communication systems for battery-free IoT: A tutorial and survey of recent research," \textit{IEEE Signal Process. Mag.}, vol. 35, no. 5, pp. 16–27, Sept. 2018.
% \bibitem{ref3} J. Jiang et al., "Long-range ambient LoRa backscatter with parallel decoding," in \textit{Proc. MobiCom}, New Orleans, Louisiana, Oct. 2021, pp. 684–696.
% \bibitem{ref4} M. Zhang et al., "Reliable backscatter with commodity BLE," in \textit{Proc. IEEE INFOCOM}, Jul. 2020, pp. 1291–1299.

% \end{thebibliography}
\bibliographystyle{IEEEtran}
\bibliography{ref}

\end{document}